\def\be{\begin{equation}}
\def\ee{\end{equation}}
\def\bea{\begin{eqnarray}}
\def\eea{\end{eqnarray}}

\documentclass[aps,prl,twocolumn,amsmath,amssymb]{revtex4}
\usepackage{epsfig,graphicx}
\usepackage{bm}

\begin{document}

%\preprint{draft}
%\draft

\title{Decoherence suppression by uncollapsing}

\author{Alexander N. Korotkov and Kyle Keane}
\affiliation{Department of Electrical Engineering and Department of
Physics and \& Astronomy, University of California, Riverside, CA
92521, USA
% \\ and Institute of Nuclear Physics, Moscow State University, Moscow 119991, Russia
}

\date{\today}

%\maketitle

\begin{abstract}
We show that the qubit decoherence due to zero-temperature energy
relaxation can be almost completely suppressed by using the quantum
uncollapsing procedure. To protect a qubit state, a partial quantum
measurement moves it towards the ground state, where it is kept
during the storage period, while the second partial measurement
restores the initial state. This procedure preferentially selects
the cases without energy decay events. Stronger decoherence
suppression requires smaller selection probability; a desired point
in this trade-off can be chosen by varying the measurement strength.
The experiment can be realized in a straightforward way using the
superconducting phase qubit.

\end{abstract}

% 03.65.Ta Foundations of quantum mechanics; measurement theory
% 03.65.Yz Decoherence; open systems; quantum statistical methods
% 03.67.Pp Quantum error correction and other methods for protection against decoherence
% 03.67.-a Quantum information
% 03.67.Lx Quantum computation architectures and implementations
% 85.25.Cp Josephson devices

\maketitle
%\narrowtext
%\vspace{1ex}
%\vspace{0.6cm}

Qubit decoherence can be efficiently suppressed via the quantum
error correction, by encoding the logical qubit in several physical
qubits and performing sufficiently frequent measurement/correction
operations \cite{Nielsen}. The use of a larger Hilbert space is also
needed in the idea of decoherence-free subspace \cite{DFS}.
  Without increasing the physical Hilbert
space, it is possible to suppress decoherence using the technique of
dynamical decoupling based on sequences of control pulses, for
example, by the ``bang-bang'' control \cite{Lloyd}. Unfortunately,
the dynamical decoupling does not help \cite{Lloyd,Pryadko-09} when
the decoherence is due to processes with short correlation
timescales, as for example for the most standard (Markovian) energy
relaxation and dephasing. The energy relaxation can in principle be
suppressed by changing properties of the qubit environment, as for
suppression of spontaneous emission in cavities \cite{Haroche};
however, this possibility does not seem very practical for
solid-state qubits. In this paper we show that the energy relaxation
in a single physical qubit can also be suppressed by using quantum
uncollapsing \cite{Kor-Jor,Katz-uncol}.

   The uncollapsing is a probabilistic reversal \cite{Kor-Jor} of a partial
quantum measurement by another measurement with an ``exactly
contradicting'' result, so that the total classical information is
zeroed, thus making possible to restore any initial quantum state.
If the second measurement gives this desired result, the initial
state is restored, while if the measurement result is different, the
uncollapsing attempt is unsuccessful. The probability of success
(selection) decreases with increasing strength of the first
measurement, so that uncollapsing has zero probability for the
traditional projective measurement. Perfect uncollapsing requires an
ideal (quantum-efficient) detector. The quantum uncollapsing has
been recently demonstrated experimentally \cite{Katz-uncol} for a
superconducting phase qubit \cite{phase-qubit}, attracting some
general interest \cite{uncol-nature}.

   The logic states in the phase qubit are represented by
two lowest energy levels in a quantum well, separated by $\sim$ 25
$\mu$eV, and the energy relaxation presents the major decoherence
process, often nearly dominating in comparison with pure dephasing
\cite{Martinis-QIP}. The experimental temperature of $\sim$ 50 mK in
this case corresponds to essentially the zero-temperature limit.
This is exactly the regime, in which uncollapsing can be used to
suppress the qubit decoherence (similar zero-temperature regime with
negligible pure dephasing is realized in transmon qubits
\cite{Houck-QIP}).

   In order to protect the qubit against
zero-temperature energy relaxation, we first apply a partial quantum
measurement (Fig.\ 1), which moves the qubit state towards the
ground state in a coherent but non-unitary way (as in
\cite{Katz-partial}). Then after the storage period we apply the
uncollapsing procedure (for the phase qubit consisting of a
$\pi$-pulse, second partial measurement, and one more $\pi$-pulse),
which restores the initial qubit state. As we see, the protocol is
very close to the existing uncollapsing experiment
\cite{Katz-uncol}. The procedure is probabilistic, since it selects
only specific results of both measurements. (In this respect it is
similar to linear optics quantum computing \cite{KLM}, which also
relies on specific measurement results.)
  If an energy relaxation event happens during the storage period,
then such case will be preferentially rejected at the selection of
the second measurement result. However, there is a trade-off: by
increasing the strength of measurements we obtain stronger
decoherence suppression, but decrease the selection probability.

\begin{figure}[tb]
  \centering
\includegraphics[width=8cm]{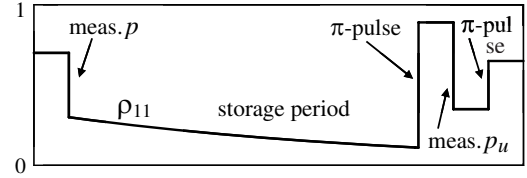}
\vspace{-0.1cm}
  \caption{Illustration of the uncollapsing sequence suppressing energy
relaxation in a phase qubit: partial measurement with strength $p$,
relatively long ``storage'' period, $\pi$-pulse, second measurement
with strength $p_u$, and $\pi$-pulse. The line illustrates evolution
of the element $\rho_{11}$ of the qubit density matrix. We select
only null-result  cases for both measurements.}
  \label{fig1}
\end{figure}

    To analyze the procedure quantitatively, let us assume that the
initial state of the qubit in the rotating frame is
$|\psi_{in}\rangle = \alpha |0\rangle +\beta |1\rangle$. The partial
measurement is performed in the standard for the phase qubit way
\cite{Martinis-QIP,Katz-partial}, by allowing the state $|1\rangle$
to tunnel out of the quantum well with the probability $p$, while
the state $|0\rangle$ cannot tunnel out. In the null-result case of
no tunneling the qubit state becomes \cite{Katz-partial,note-phase}
\vspace{-0.1cm}
    \begin{equation}
    |\psi_1\rangle = \alpha_1 |0\rangle +\beta_1 |1\rangle =
 \frac{\alpha |0\rangle +\beta \sqrt{1-p} \, |1\rangle }
{\sqrt{|\alpha|^2+|\beta|^2 (1-p)} },
    \label{part-meas}\end{equation}
and the probability of no tunneling is $P_1 = |\alpha|^2+|\beta|^2
(1-p)$.
    After the storage period $\tau$ the qubit state is no longer
pure because of (zero-temperature) energy relaxation with the rate
$\Gamma =1/T_1$. However, it is technically easier for us to
``unravel'' this process into ``jump'' and ``no jump'' scenarios,
and work with pure states (this is a purely mathematical trick,
which does not assume any jumps in reality). So, we can think that
after the storage time $\tau$ the qubit jumps into the state
$|0\rangle$ with the total probability $P_2^{|0\rangle}=P_1
|\beta_1|^2 (1-e^{-\Gamma \tau})$, while it ends up in the state
   \vspace{-0.2cm}
    \begin{equation}
    |\psi_2\rangle = \alpha_2 |0\rangle +\beta_2 |1\rangle =
 \frac{\alpha |0\rangle +\beta \sqrt{1-p} \, e^{-\Gamma \tau/2} |1\rangle }
{\sqrt{|\alpha|^2+|\beta|^2 (1-p) \, e^{-\Gamma \tau}} }
    \label{after-tau}\end{equation}
with ``no jump'' probability $P_2^{\rm nj}=|\alpha|^2+|\beta|^2
(1-p) \, e^{-\Gamma \tau}$. Notice that we made the Bayesian-like
update \cite{Kor-99} of the qubit state $|\psi_2\rangle$ in the ``no
energy jump'' scenario; such update must be done even when the jump
is not monitored, as can be easily checked by comparing the
resulting density matrices. Also notice that the denominator in Eq.\
(\ref{after-tau}) is $(P_2^{\rm nj})^{1/2}$, as expected from the
general theory of quantum measurement \cite{Nielsen}.

    After applying the $\pi$-pulse the qubit state becomes either
$|1\rangle$ or $\alpha_3 |0\rangle +\beta_3 |1\rangle=\alpha_2
|1\rangle +\beta_2 |0\rangle$ with the same probabilities
$P_2^{|0\rangle}$ and $P_2^{\rm nj}$. Then after the second
(uncollapsing) measurement with strength $p_u$, in the no-tunneling
case the qubit remains in the state $|1\rangle$ with the total
probability $P_4^{|1\rangle}=P_{2}^{|0\rangle}(1-p_u)$, while its
state becomes
   \vspace{-0.2cm}
    \begin{equation}
    \alpha_4 |0\rangle +\beta_4 |1\rangle =
 \frac{\beta \sqrt{1-p} \, e^{-\Gamma \tau/2}  |0\rangle +\alpha \sqrt{1-p_u} \,  |1\rangle }
{\sqrt{|\alpha|^2 (1-p_u)+|\beta|^2 (1-p) \, e^{-\Gamma \tau}} }
    \label{after-second-meas}\end{equation}
with probability $P_4^{\rm nj}=|\alpha|^2 (1-p_u)+|\beta|^2 (1-p) \,
e^{-\Gamma \tau}$. Finally, the second $\pi$-pulse produces either
the state $|0\rangle$ with probability
$P_{f}^{|0\rangle}=P_4^{|1\rangle}$ or the final state
$|\psi_{f}\rangle = \beta_4|0\rangle +\alpha_4 |1\rangle$ with
probability $P_{f}^{\rm nj}=P_4^{\rm nj}$.

    It is easy to see that in the ``no jump'' scenario the best
(exact) restoration of the initial state is when $p_u=1-e^{-\Gamma
\tau}(1-p)$, and in this case the final state is
\begin{eqnarray}
&& \hspace{-0.4cm}  |\psi_{f}\rangle=  |\psi_{in}\rangle \,\,\, {\rm
with\,\, probability}
  \,\,\,
  P_{f}^{\rm nj}= (1-p) \, e^{-\Gamma \tau} , \quad
    \label{final-nj}    \\
&& \hspace{-0.4cm} |\psi_{f}\rangle=  |0\rangle \,\,\, {\rm with}
\,\,\,
  P_{f}^{|0\rangle}= |\beta |^2 (1-p)^2 e^{-\Gamma
  \tau} (1-e^{-\Gamma \tau}) . \quad
    \label{final-0}\end{eqnarray}
In the language of density matrix this means that both measurements
produce null results (no tunneling) with the selection probability
$P_f=P_f^{\rm nj}+P_f^{|0\rangle}$, and in such a case the final
qubit state is
    \begin{equation}
\rho_f = \left( P_f^{\rm nj} |\psi_{in}\rangle \langle \psi_{in} | +
P_f^{|0\rangle } |0\rangle  \langle 0 | \right) /(P_f^{\rm nj}+
P_f^{|0\rangle }).
    \label{rho-final}\end{equation}

    An important observation is that the ``good'' probability
$P_f^{\rm nj}$ scales as $1-p$ with the measurement strength $p$,
while the ``bad'' probability $P_f^{|0\rangle}$ scales as $(1-p)^2$.
Therefore, choosing $p$ close to 1, we can make the final qubit
state {\it arbitrarily close to the initial state}, even in the
presence of a significant decoherence due to energy relaxation
($\Gamma \tau \agt 1$). This is the main result of our paper.

    It is tempting to say that the decoherence is suppressed because
the storage state is close to the ground state, where the energy
relaxation is naturally suppressed.
     However, a better explanation of the effect is
that for the basis state $|0\rangle$ the energy relaxation is absent
by itself, while for the basis state $|1\rangle$ the mechanism is
the following: the first measurement keeps it as $|1\rangle$, but if
the state jumps down to $|0\rangle$ during the storage period, then
most likely there will be tunneling during the second measurement,
and therefore such events will be eliminated by the selection of
only null-result cases.

    We can characterize the performance of the procedure by
calculating the fidelity of the quantum state storage and analyzing
its increase with the measurement strength $p$. The fidelity of a
quantum operation is usually defined as $F_\chi=\mbox{Tr} (\chi
\chi_0)$ where $\chi$ is the quantum process tomography (QPT) matrix
\cite{Nielsen}, while $\chi_0$ is the QPT matrix of the desired
unitary operation (which in our case is the identity mapping). In
particular, this characteristic has been used in the QPT experiments
which involve selection of certain measurement results
\cite{CNOT-optics,Katz-uncol}, even though strictly speaking it is
inapplicable in this case. The reason for the inapplicability is
that the QPT approach assumes a linear quantum operation, while the
selection procedure involves renormalization of the density matrix,
which in general makes the mapping nonlinear. Nevertheless, as
discussed below, in our case the fidelity $F_\chi$ can still be
defined in a ``naive'' way by using 4 standard initial qubit states
to calculate $\chi$ (as was done in \cite{Katz-uncol}), and the
result practically coincides with another, more rigorous definition.
The definition which still works in the presence of selection is the
average state fidelity \cite{Nielsen} $F_{av} = \int {\mbox Tr}
(\rho_f U_0|\psi_{in} \rangle \langle \psi_{in} | )\, d|\psi_{in}
\rangle $, where $U_0=\openone$ is the desired unitary operator,
$\rho_f(|\psi_{in}\rangle)$ is the actual mapping [given by Eq.\
(\ref{rho-final})], and the normalized integral is over all pure
initial states $|\psi_{in}\rangle$. For trace-preserving operations
(without selection) $F_{av}=(F_\chi d+1)/(d+1)$ \cite{Nielsen-02},
where $d=2$ is the dimension of our Hilbert space. Therefore, it is
natural to define a scaled average fidelity $F_{av}^s\equiv (3
F_{av}-1)/2$, which would coincide with $F_\chi$ in a no-selection
case.

    The state fidelity $F_{st}={\mbox Tr} (\rho_f |\psi_{in} \rangle \langle \psi_{in} |
)$ between the desired unevolved state $|\psi_{in}\rangle$ and the
actual state $\rho_f$ given by Eq.\ (\ref{rho-final}) is
%    \begin{equation}
$F_{st}=1-|\beta|^2P_f^{|0\rangle}/P_f$.
%    \label{F-st}\end{equation}
    In order to average $F_{st}$ over the initial state
we use the integration result
    \begin{equation}
    \left\langle \frac{|\beta|^4}{A+B|\beta|^2} \right\rangle_{\rm Bl} =
    \frac{1}{2B}-\frac{A}{B^2} +\frac{A^2}{B^3} \, \ln
    (1+\frac{B}{A}),
    \label{integral-1}\end{equation}
where $\langle .. \rangle_{\rm Bl}$ denotes averaging over the Bloch
sphere. Using $A=1$ and $B=(1-p)(1-e^{-\Gamma \tau})$ [see Eqs.\
(\ref{final-nj})--(\ref{rho-final}), the common factor
$(1-p)e^{-\Gamma\tau}$ is canceled], we thus find
    \begin{equation}
    F_{av}=\frac{1}{2}+\frac{1}{C}-\frac{\ln (1+C)}{C^2}, \,\,\,
    C=(1-p)(1-e^{-\Gamma\tau}) ,
    \label{F-av}\end{equation}
and the corresponding scaled fidelity $F_{av}^s=(3 F_{av}-1)/2$.
    It is important to notice that while the fidelity $F_{av}^s$
increases with the measurement strength $p$, this happens for the
price of decreasing the average selection probability $\langle
P_f\rangle_{Bl}=(1-p)e^{-\Gamma\tau} (1+C/2)$. In particular, for
$p\rightarrow 1$ we have $F_{av}^s\rightarrow 1$, but $\langle
P_f\rangle_{Bl}\rightarrow 0$.

    In experiments the one-qubit process fidelity $F_\chi$ is usually
defined by starting with four specific initial states: $|0\rangle$,
$|1\rangle$, $(|0\rangle +|1\rangle)/\sqrt{2}$, and $(|0\rangle
+i|1\rangle)/\sqrt{2}$, measuring the corresponding final states
$\rho_f$, then calculating the $\chi$-matrix, and finally obtaining
$F_\chi$. Even for a non-linear quantum operation this is a
well-defined procedure (just the result may depend on the choice of
the four initial states), so it is meaningful to calculate $F_\chi$
defined in this (naive) way. It is obvious that such defined
$F_\chi$ coincides with $F_\chi$ for a linear trace-preserving
operation, which gives the same final states for the four chosen
initial states. Next, we use the fact \cite{Nielsen-02} that the
average fidelity $\tilde{F}_{av}$ for this ``substitute'' operation
is equal to $F_{st}$ averaged over only 6 initial states:
$|0\rangle$, $|1\rangle$, $(|0\rangle \pm |1\rangle)/\sqrt{2}$, and
$(|0\rangle \pm i|1\rangle)/\sqrt{2}$. Since in our case $F_{st}$
   %given by Eq.\(\ref{F-st})
   is phase-insensitive, we get $\tilde{F}_{av}=[F_{st}(|0\rangle)+F_{st}(|1\rangle)+
4F_{st}(\frac{|0\rangle+|1\rangle}{\sqrt{2}})]/6$, which gives
    \begin{equation}
    \tilde{F}_{av}= \frac{1}{6}+\frac{1}{6(1+C)}
    +\frac{4+C}{3(2+C)} .
    \label{F-av-naive}\end{equation}
Then the ``naive'' fidelity is simply
$F_\chi=(3\tilde{F}_{av}-1)/2$.

\begin{figure}[tb]
  \centering
\includegraphics[width=8cm]{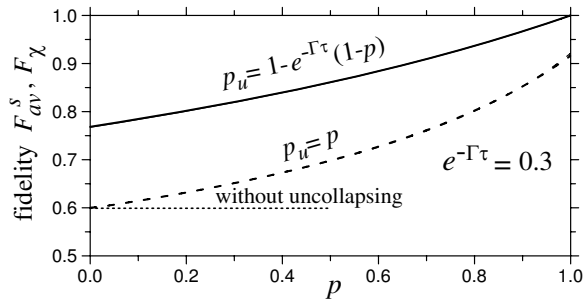}
\vspace{-0.1cm}
  \caption{Fidelity of the quantum state storage using uncollapsing, as a function
of the first measurement strength $p$  for $e^{-\Gamma\tau}=0.3$.
Thick lines show $F_{av}^s$, while thin lines (practically
indistinguishable from thick lines) show $F_\chi$. Solid and dashed
lines are for two choices of the second measurement strength $p_u$.
Horizontal dotted line indicates fidelity without uncollapsing. }
  \label{fig2}
\end{figure}

    Efficiency of the energy relaxation suppression by uncollapsing
is illustrated in Fig.\ 2 by plotting (solid lines) the scaled
average fidelity $F_{av}^s$ and the ``naive'' fidelity $F_\chi$ as
functions of the measurement strength $p$ for a quite significant
energy relaxation: $e^{-\Gamma t}=0.3$. Notice that $F_{av}^s$ and
$F_\chi$ are practically indistinguishable (within thickness of the
lines), despite different functional dependences in Eqs.\
(\ref{F-av}) and (\ref{F-av-naive}). Also notice that even for $p=0$
the fidelities differ from the fidelity without uncollapsing
($F_\chi =1/2+e^{-\Gamma \tau}/4+e^{-\Gamma\tau /2}/2\approx 0.6$),
shown by the dotted line in Fig.\ 2. This is because we assumed
$p_u=1-e^{-\Gamma\tau}(1-p)$, so $p_u\neq 0$ even for $p=0$, and the
second measurement improves the fidelity. If we choose $p_u=p$
(dashed lines) as in the standard uncollapsing
\cite{Kor-Jor,Katz-uncol}, then the case $p=0$ is equivalent to the
absence of any procedure. [The dashed lines are calculated in a
similar way, assuming $p_u=p$ in Eq.\ (\ref{after-second-meas}).]
    It is interesting to notice that if we numerically maximize the
fidelity $F_{av}^s$ by optimizing over $p_u$, then we can get larger
$F_{av}^s$ (for the same $p$) than in the case
$p_u=1-e^{-\Gamma\tau}(1-p)$; however, this will decrease the
selection probability $\langle P_f\rangle_{Bl}$, and for the same
$\langle P_f\rangle_{Bl}$ such optimization slightly decreases
$F_{av}^s$.

    So far we assumed that the energy relaxation happens only during
the storage period, while there is no decoherence during the
uncollapsing procedure (measurements and $\pi$-pulses). Even though
such assumption is justified since the storage period for a quantum
memory is supposed to be relatively long, let us take a step closer
to reality and take into account energy relaxation during all
durations illustrated by horizontal lines in Fig.\ 1 (except the
last one, which is after the procedure is finished). The energy
relaxation (still zero-temperature) will be characterized by
parameters $\kappa_i=\exp(-\Gamma \tau_i)$, $i=1$--4, where $\tau_1$
is the duration before the first measurement, $\tau_2=\tau$ is the
storage period, $\tau_3$ is the duration between the first
$\pi$-pulse and second measurement, and $\tau_4$ is between the
second measurement and second $\pi$-pulse (the measurements and
$\pi$-pulses are still assumed ideal). Using the same derivation as
above and selecting only the null-result cases for both
measurements, we can show that for the initial state
$|\psi_{in}\rangle =\alpha |0\rangle +\beta |1\rangle$ the final
state can be unraveled as
    \be
    |\psi_{f}^{\rm nj}\rangle = \frac{\alpha \sqrt{\kappa_3\kappa_4 (1-p_u)} |0\rangle +
\beta \sqrt{\kappa_1\kappa_2 (1-p)} |1\rangle }{(P_f^{\rm
nj})^{1/2}}
    \label{psi-f}\ee
with the ``no jump'' probability $P_f^{\rm nj}= |\alpha|^2
\kappa_3\kappa_4 (1-p_u)+ |\beta|^2 \kappa_1\kappa_2 (1-p)$, the
state $ |\psi_{f}\rangle =|0\rangle$ with probability
$P_f^{|0\rangle}=|\alpha|^2
[1-\kappa_3+\kappa_3(1-p_u)(1-\kappa_4)]+|\beta|^2[1-\kappa_1+
\kappa_1(1-p)(1-\kappa_2)][1-\kappa_3+\kappa_3(1-p_u)(1-\kappa_4)]$,
and also $|\psi_{f}\rangle =|1\rangle$ with probability
$P_f^{|1\rangle}=|\beta|^2[1-\kappa_1+
\kappa_1(1-p)(1-\kappa_2)]\kappa_3(1-p_u)\kappa_4]$ (all terms in
these formulas have rather obvious physical meaning). Actual density
matrix is then $\rho_f =(P_f^{\rm nj}|\psi_f^{\rm nj}\rangle \langle
\psi_f^{\rm nj}| + P_f^{|0\rangle} |0\rangle \langle 0| +
P_f^{|1\rangle} |1\rangle \langle 1| )/(P_f^{\rm
nj}+P_f^{|0\rangle}+P_f^{|1\rangle})$ and the selection probability
is $P_f=P_f^{\rm nj}+P_f^{|0\rangle}+P_f^{|1\rangle}$. It is also
rather simple to take into account the additional decoherence due to
the pure dephasing with rate $\Gamma_\varphi$. It can be shown that
the only change will be the pure dephasing of the state
(\ref{psi-f}) with the factor $\kappa_\varphi=\exp(-\Gamma_\varphi
\sum_{i=1}^4 \tau_i)$.

    The state fidelity then can be calculated in a straightforward
way, and the averaging over the initial state can be performed as
above using the integration result (\ref{integral-1}) and similar
result $\langle |\alpha|^4/(A+B|\beta|^2)\rangle_{Bl} =
-(3/2B)-(A/B^2)+(1/B)(1+B/A)^2\ln (1+B/A)$. The final result for the
scaled averaged fidelity $F_{av}^s$ is analytical, but rather
lengthy (as well as for $F_\chi$ and $\langle P_f\rangle_{Bl}$).

\begin{figure}[tb]
  \centering
\includegraphics[width=8cm]{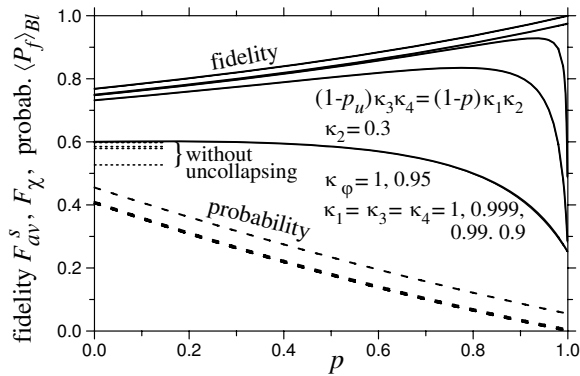}
\vspace{-0.1cm}
  \caption{Solid lines: fidelities $F_{av}^s$ and $F_\chi$  of the
state storage (still practically indistinguishable from each other),
taking into account the energy relaxation and pure dephasing during
all parts of the uncollapsing procedure, for several sets of
parameters (see text). Dashed lines: corresponding selection
probabilities $\langle P_f\rangle_{Bl}$ (reverse order of curves).
Dotted lines: corresponding fidelities without uncollapsing
($p=p_u=0$). }
  \label{fig3}
\end{figure}

    Solid lines in Fig.\ 3 show the $p$-dependence of the fidelities
$F_{av}^s$ and $F_\chi$ (they are still indistinguishable, being
within the thickness of the line), for which we choose $p_u$ from
equation ${\kappa}_3{\kappa}_4(1-p_u)={\kappa}_1{\kappa}_2 (1-p)$
which comes from Eq.\ (\ref{psi-f}) and generalizes the equation
$1-p_u=e^{-\Gamma\tau} (1-p)$.
 For all solid lines we
assume $k_2=0.3$. The upper line is for the ideal case
${\kappa}_1={\kappa}_3={\kappa}_4={\kappa}_\varphi =1$ (so it is the
same as in Fig.\ 2). For all other lines ${\kappa}_\varphi =0.95$,
while ${\kappa}_1={\kappa}_3={\kappa}_4=1$, 0.999, 0.99, 0.9 (from
top to bottom). Dotted lines show corresponding fidelities in
absence of uncollapsing ($p=p_u=0$; then $F_{av}^s=F_\chi
=1/4+\kappa_E/4+\kappa_\varphi \sqrt{\kappa_{E}}/2$, where
$\kappa_E=\kappa_1\kappa_2\kappa_3\kappa_4$). The dashed lines show
the selection probability $\langle P_f\rangle$ of the procedure;
these lines go in the opposite sequence (from bottom to top)
compared to the solid and dotted lines.

    As we see from Fig. 3, the uncollapsing essentially does not affect decoherence
due to the pure dephasing ($\kappa_\varphi$), while the energy
relaxation during the elements of the procedure ($\kappa_1,
\kappa_3,\kappa_4$) has a less trivial effect: for small $p$ it just
reduces the fidelity, while for $p\rightarrow 1$ it causes fidelity
to drop down to 0.25 (this value corresponds to complete
decoherence; the fidelity decrease is mainly affected by
$\kappa_3$). Notice that the lowest solid line does not show a
noticeable increase of the fidelity with $p$ before it starts to
decrease. This behavior is similar to the results of the
uncollapsing experiment \cite{Katz-uncol}, in which the ``storage''
time between the first measurement and $\pi$-pulse was not longer
than other durations. Changing the experimental protocol of
\cite{Katz-uncol} by relative increase of the storage time, we would
expect to observe initial increase of the fidelity with $p$, thus
confirming that uncollapsing can suppress decoherence.

    Notice that all solid lines in Fig.\ 3 are significantly above the
standard fidelity (dotted lines, $p=p_u=0$) for moderate measurement
strength $p$. Significant increase of the fidelity is especially
remarkable in view of the fact \cite{Pryadko-09} that arbitrary
Hamiltonian evolution cannot even slightly improve the fidelity in
our case. So, the uncollapsing (which involves selection of certain
measurement results) is {\it the only} known to us way of improving
the qubit storage fidelity against energy relaxation, which does not
rely on encoding a logical qubit in a larger Hilbert space. Our idea
also works for entangled qubits \cite{note-entangled}.

We thank Leonid Pryadko and Alexei Kitaev for fruitful discussions.
 This work was supported by NSA and IARPA under ARO grant
W911NF-08-1-0336.

\vspace{-0.4cm}

%\begin{references}


\begin{thebibliography}{99}
\bibitem{note-phase} In real experiment \protect\cite{Katz-partial}
the partial measurement leads to an additional phase shift between
states $|0\rangle$ and $|1\rangle$. We neglect it in Eq.\
(\ref{part-meas}) because it can be easily compensated and sometimes
cancels out automatically \protect\cite{Katz-uncol}.
\bibitem{note-entangled} Suppose now that the qubit is entangled
with other qubits, which do not decohere. Parameterizing initial
state as $\alpha |0\rangle |\psi_0\rangle +\beta |1\rangle
|\psi_1\rangle$, it is easy to show that Eq.\ (\protect\ref{psi-f})
changes only trivially, and the contribution to the state fidelity
from the ``no jump'' scenario does not change at all, while several
terms in the contribution from $|0\rangle$ and $|1\rangle$ will be
multiplied by $|\langle \psi_0|\psi_1\rangle|^2$. Therefore, the
results of this paper do not change qualitatively for entangled
qubits, though they change quantitatively.
\end{thebibliography}
\end{document}